\documentclass[a4paper, oneside, twocolumn, notitlepage, 10pt]{extarticle_ecoc}
\usepackage{ecoc}
\usepackage[nolist]{acronym}

\acrodef{AI}{artificial intelligence}
\acrodef{API}{application programming interface}
\acrodef{ASE}{amplifier spontaneous emission}
\acrodef{CD}{chromatic dispersion}
\acrodef{DT}{digital twin}
\acrodef{DSP}{digital signal processing}
\acrodef{EDFA}{Erbium-doped fiber amplifier}
\acrodef{EEPN}{equalization-enhanced phase noise}
\acrodef{FEC}{forward error correction}
\acrodef{GN}{Gaussian noise}
\acrodef{gNMI}{gRPC network management interface}
\acrodef{gRPC}{gRPC remote procedure call}
\acrodef{GSNR}{generalized signal-to-noise ratio}
\acrodef{ML}{machine learning}
\acrodef{ODC}{optical domain controller}
\acrodef{ONF}{Open Networking Foundation}
\acrodef{OPM}{optical performance monitoring}
\acrodef{OSNR}{optical signal-to-noise ratio}
\acrodef{PDL}{polarization-dependent loss}
\acrodef{PMD}{polarization-mode dispersion}
\acrodef{QoT}{quality of transmission}
\acrodef{REST}{representational state transfer}
\acrodef{RMSA}{routing, modulation, and spectrum assignment}
\acrodef{ROADM}{reconfigurable optical add-drop multiplexer}
\acrodef{SDN}{software-defined networking}
\acrodef{SNR}{signal-to-noise ratio}
\acrodef{T-API}{Transport-API}

\begin{document}
\selectlanguage{english}    % Standard Language

%-------------------------------------------------- Title -----------------------------------------------------%

\title{TwinLight: A Multi-Timescale Physical-Layer-Aware T-API-Compliant Optical Domain Controller Digital Twin}%

%------------------------------------------------- Authors-----------------------------------------------------%

\author{
    Carlos Natalino\textsuperscript{(1)}, Piotr Lechowicz\textsuperscript{(1)}, Renato Ambrosone\textsuperscript{(2)},\\
    Enrico Miotto\textsuperscript{(2)}, Vittorio Curri\textsuperscript{(2)}, Paolo Monti\textsuperscript{(1)}
}

\maketitle                  % Create title and author

%------------------------------------------ Description of Authors ----------------------------------------------%

\begin{strip}
    \begin{author_descr}

        \textsuperscript{(1)} Department of Electrical Engineering, Chalmers University of Technology, Gothenburg, Sweden.
        \textcolor{blue}{\uline{carlos.natalino@chalmers.se}}

        \textsuperscript{(2)} Politecnico di Torino, Torino, Italy.

    \end{author_descr}
\end{strip}

% \setstretch{1.1}
%-------------------------------------------------- Footnote -------------------------------------------------------%
\renewcommand\footnotemark{}
\renewcommand\footnoterule{}
%\let\thefootnote\relax\footnotetext{text}

%-------------------------------------------------- Abstract ---------------------------------------------------------%

\begin{strip}
    \begin{ecoc_abstract}
        % NOTE: Don't use a blank line here but start abstract right away to avoid an extra line break -- 35 words for the demos!
        We demonstrate TwinLight, an open-source optical-network digital twin acting as an optical domain controller. Its T-API REST and telemetry streaming interfaces expose a physical-layer engine that combines a GNPy steady-state baseline with four analytical transients. \copyright2026 The Author(s)
    \end{ecoc_abstract}
\end{strip}

%-------------------------------------------------- Introduction Section -------------------------------------------------------%

\section{Review}

Disaggregated and software-defined optical networks increasingly rely on \acp{DT} to de-risk control-plane decisions\footnote{ECOC 2026 paper ID Tu2-Ex1/1440.}\footnote{The code of the tool described in this paper is available at \url{https://github.com/carlosnatalino/TwinLight}.}.
A faithful software replica of the optical domain helps validate \ac{SDN} control logic, explore what-if scenarios, prototype \ac{RMSA} algorithms, and generate training data for \ac{ML} models \cite{Vilalta_2023_ApplyingDigitalTwins, Wang_2024_DigitalTwinOptical}.
This requirement becomes more relevant as autonomous and agentic strategies begin to operate in closed loop, observe physics-consistent consequences, and be validated against the same northbound interfaces that a production controller would expose \cite{Etezadi_2023_DemonstrationDRLbasedIntelligent, Song_2025_SynergisticInterplayLarge}.

The tooling landscape that has emerged to meet these needs resolves into three disjoint families of solutions \cite{Deshpande_2023_ComparativeStudyNetwork}.
First, physical-layer tools such as GNPy compute steady-state \ac{QoT} with high fidelity \cite{Ferrari_2020_GNPyOpenSource, Curri_2022_GNPyModelPhysical}, but are not designed to expose the persistent state and standardized northbound interfaces that a controller-facing \ac{ODC} must provide.
Second, network emulators such as Mininet-Optical capture topology, embed \ac{OPM}, and expose NETCONF and gNMI interfaces aligned with OpenConfig and OpenROADM device models, yet operate exclusively in steady state \cite{Lantz_2020_DemonstrationSoftwareDefinedPacketOptical}.
Within the same family, data-driven digital twins such as OCATA refine steady-state \ac{QoT} estimation through telemetry-driven model tuning, but expose ad-hoc northbound interfaces and likewise lack an analytical model of temporal impairments \cite{Gonzalez_2025_DemonstratingLightpathOperation}.
A third family combines physical-layer-aware \ac{QoT} with \ac{RMSA} logic and recalculates mutual impact when a new lightpath perturbs the \ac{SNR} of existing services \cite{Natalino_2024_OpticalNetworkingGym, Garbhapu_2023_PhysicallayerawareNetworkSimulator}, but lacks both a standards-compliant northbound interface and analytical models of temporal physical impairments.
Considered in isolation, none of these families can serve as an \ac{ODC} for a production orchestrator.

Two further developments sharpen the requirements that a controller-facing \ac{DT} must satisfy.
The first concerns physics, including physical effects that span multiple timescales.
The community has converged on the \ac{GN} family of models as the working description of nonlinear interference in coherent transmission, and on GNPy as its \emph{de facto} implementation, with recent field validation against partially-disaggregated multi-vendor deployments \cite{Carena_2014_EGNModelNonlinear, Ferrari_2020_GNPyOpenSource, Borraccini_2023_ExperimentalDemonstrationPartially}.
Real coherent links, however, undergo impairments that no steady-state snapshot can capture.
Channel add-drop events trigger microsecond-scale power excursions in cascaded \acp{EDFA} \cite{Bononi_1998_DopedfiberAmplifierDynamics}.
The state of polarisation drifts on second-to-day timescales \cite{Gordon_2000_PMDFundamentalsPolarization}, and \ac{PDL} adds a stochastic \ac{OSNR} penalty over the same range \cite{Zarkosvky_2020_StatisticalDistributionPolarizationdependent, DAmico_2023_ExperimentalProbingModeling, Miotto_2025_StatisticalAssessmentSystem}.
Accumulated dispersion couples with the local-oscillator linewidth to produce \ac{EEPN} on the minute scale \cite{Shieh_2008_EqualizationenhancedPhaseNoise}, while diurnal thermal cycles drift the \ac{CD} and modulate span loss \cite{Kato_2000_TemperatureDependenceChromatic, Berkdemir_2005_TemperatureDependentPerformance}.
Together, these mechanisms span nine orders of magnitude in time.

The second concerns the \acp{API} that the \ac{DT} exposes.
The \ac{ONF} \ac{T-API} defines a \ac{REST} interface that an \ac{ODC} must expose to a hierarchical orchestrator, which may integrate optical services with other technology domains such as packet-switching and wireless networks.
\ac{T-API} exposes topology, connectivity, photonic-media, and path-computation contexts that allow comprehensive control of the infrastructure associated with the 
\ac{ODC} \cite{Manso_2021_TAPIenabledSDNControl, OpenNetworkModelsandInterfacesONMI__OnfspecificationsTR547_TAPIReferenceImplementationAgreementMain}.
\Ac{T-API} streaming, aligned with the OpenConfig \ac{gNMI} subscription model, has emerged as the telemetry counterpart \cite{Karunakaran_2024_TAPIbasedTelemetryStreaming}.
A \ac{DT} that aims to operate under a production orchestrator must therefore implement both protocols natively.

\section{Novelty}

\begin{figure}[!t]
    \centering
    \includegraphics[width=0.95\columnwidth]{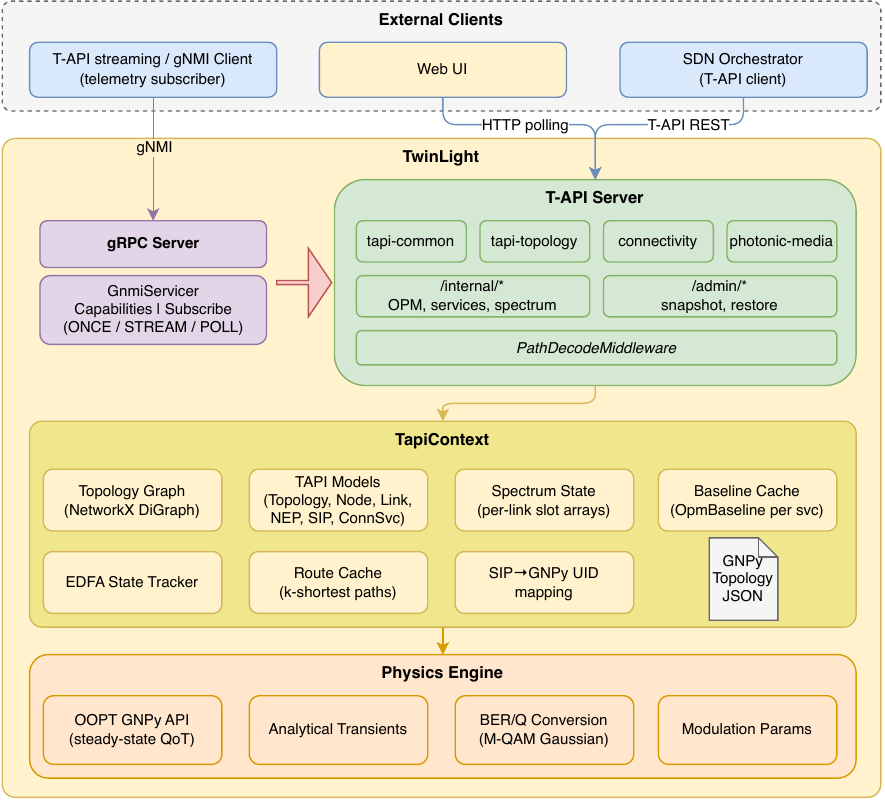}
    \caption{TwinLight architecture.}
    \label{fig_twinlight_architecture}
\end{figure}

The overall architecture of TwinLight is shown in Fig.~\ref{fig_twinlight_architecture}.
Northbound, the \ac{DT} exposes two interfaces:
\emph{(i)} a \ac{T-API} \ac{REST} interface that can be used to interface with orchestrators and
\emph{(ii)} a \ac{T-API} streaming interface to telemetry subscribers.
The \ac{DT} also provides a built-in web user interface that uses the same \ac{T-API} \ac{REST} surface, as well as internal endpoints for administrative tasks.
A single stateful context manages the topology graph, the connectivity services, the per-link spectrum state, and the per-service \ac{QoT} baselines, so that physics, state, and standard interfaces remain consistent without inter-process synchronization.
The physics engine combines a GNPy steady-state baseline with the four analytical transient models and a modulation-format-aware bit-error-rate conversion, feeding back into the same context that the northbound interfaces serve.

TwinLight's novelty lies in three compounding contributions.
First, TwinLight is a \ac{DT} that serves as a complete \ac{ODC}, simultaneously exposing \ac{T-API} and 
telemetry streaming interfaces over a live physics engine, with both steady-state \ac{QoT} and transients.
The same physics engine is exposed through \ac{T-API} and the built-in web user interface, such that control logic, telemetry consumers, and human operators share a single source of truth.

Second, TwinLight provides a \ac{PDL} treatment appropriate for coherent receivers.
The model cascades a per-hinge polarization transfer across \ac{ROADM} and \ac{EDFA} hinges of each lightpath following the hinge formulation of \cite{Zarkosvky_2020_StatisticalDistributionPolarizationdependent}.
Each hinge is assigned a Maxwell-distributed magnitude \cite{Mecozzi_2002_StatisticsPolarizationdependentLoss} seeded deterministically from its identifier, and its alignment drifts with a slightly incommensurate period, so that the joint trajectory becomes quasi-periodic and ergodically fills the joint \ac{PDL} ensemble without an explicit Monte-Carlo pass.
The accumulated \ac{PDL} is then mapped to a coherent-receiver \ac{OSNR} penalty using the experimentally characterized treatment of \cite{DAmico_2023_ExperimentalProbingModeling, Miotto_2025_StatisticalAssessmentSystem}.
The mapping is sensitive to a configurable \ac{ASE}-distribution scenario that reflects the transmitter-, distributed-, and receiver-side cases reported in the same literature.

Third, TwinLight composes four transient impairment models onto the GNPy steady-state baseline through a \ac{QoT} combiner that maintains \ac{GSNR} and \ac{OSNR} as independent metrics and routes each impairment to its physically correct counterpart.
\Ac{EDFA} reservoir dynamics, polarisation effects, and the thermal loss component of the environmental drift act on both \ac{GSNR} and \ac{OSNR}.
\Ac{EEPN}, a \ac{DSP}-domain impairment, acts only on \ac{GSNR}.
Thermal \ac{CD} drift updates the reported \ac{CD} metric directly and modulates \ac{GSNR} indirectly through the \ac{EEPN} term, the only path by which a \ac{DSP}-compensated dispersion change reaches a coherent receiver.
The combiner produces a six-metric \ac{OPM} tuple at every time instance, comprising \ac{OSNR}, \ac{GSNR}, pre-\ac{FEC} BER, Q-factor, \ac{CD}, and \ac{PMD}, on top of a baseline that respects the physical separation between linear, nonlinear, polarisation, phase-noise, and environmental contributions.
From the same combined state, the combiner also synthesizes a constellation diagram and an eye diagram.
The diagrams are parameterized by the \ac{GSNR}, the combined transmitter and local-oscillator linewidth, and the \ac{PMD}, following the Gaussian-mixture representation introduced by OCATA \cite{Sequeira_2023_OCATADeeplearningbasedDigital, Gonzalez_2025_DemonstratingLightpathOperation} and the standard \ac{PMD}-driven jitter mapping \cite{Gordon_2000_PMDFundamentalsPolarization}.
The artefact thereby exposes physics-consistent observability at three abstraction levels: aggregated metrics, scatter statistics, and time-domain waveforms, all derived from the same physics state.

\section{ECOC Relevance}

Network operators and vendors of \ac{SDN} orchestrators repeatedly face the problem of integrating new control logic against a realistic, standards-compliant \ac{ODC}.
Such integration requires access to laboratory hardware that is expensive, scarce, and routinely committed to production-readiness validation rather than open experimentation.
A free, scriptable \ac{DT} that exposes the same northbound interfaces as a real \ac{ODC} removes that bottleneck and lets integration work start before any optics are powered on.
Academic researchers working on \ac{RMSA}, control-plane algorithms, agentic \ac{AI}, \ac{ML}-based \ac{QoT} estimation, and failure detection and prediction gain a reproducible physical-layer baseline that can generate realistic training data on demand.
The broader optical-networking community, which has spent the last decade converging on the \ac{GN} model and GNPy as the open steady-state physics baseline \cite{Ferrari_2020_GNPyOpenSource, Curri_2022_GNPyModelPhysical} and on \ac{T-API} as the open northbound interface \cite{Manso_2021_TAPIenabledSDNControl, Karunakaran_2024_TAPIbasedTelemetryStreaming}, gains a concrete embodiment of that convergence in a single artifact.
TwinLight is best suited for a live demonstration because its value proposition is fundamentally aligned with the Demo Zone format.
Visitors can inspect how \ac{T-API} works in real time, establish and decommission lightpaths, change physical-layer parameters, and observe the cascading effects.

\section{Demo Implementation}

The demo is built to convince a visitor of three things, in order.
A \ac{T-API} provisioning call against the \ac{DT} triggers \ac{QoT}-aware \ac{RMSA} that behaves the way an \ac{ODC} \ac{RMSA} module would.
A telemetry subscription on the resulting service streams real-time \ac{OPM} the way an operational telemetry pipeline would.
Operator-level events produce physics-consistent time-varying \ac{OPM} that can be visually correlated with the underlying transient model on screen.

Two pre-loaded topologies are selectable from the user interface.
A small one supports fast iteration and easy visualization, while a continental-scale topology with tens of \acp{ROADM} and roughly two hundred fiber spans \cite{Ferrari_2020_GNPyOpenSource} supports realistic experimentation at scale.
Both run in the C-band on an 88-channel grid with 50 GHz spacing, with three coherent modulation formats selectable per service, following the standard GNPy implementation.
Snapshot and restore endpoints instantly reset state between visitors, so a full demonstration cycle remains manageable even with continuous traffic.

The demo setup is intentionally minimal.
A poster contains an overview of the demonstration.
A single (presenter's) laptop hosts the entire \ac{DT}, including its web interface.
An external monitor displays the web interface in a large format, while the laptop screen shows a presentation with a few slides covering the architecture and theory behind TwinLight.
There is no optical hardware in the demo.

\begin{figure}[!t]
    \centering
    \includegraphics[width=0.9\columnwidth]{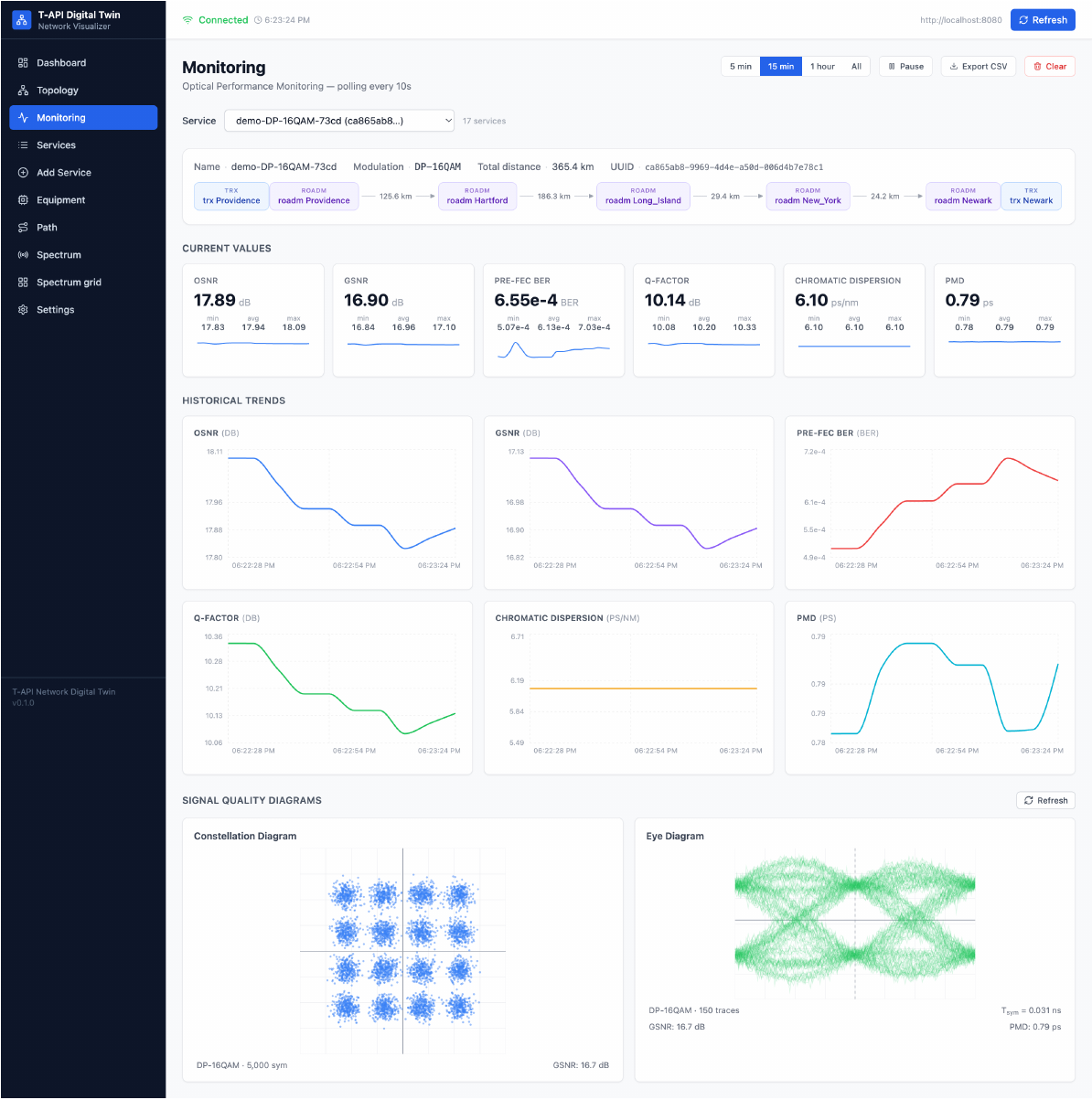}
    \caption{TwinLight preliminary web interface.}
    \label{fig_twinlight_web_interface}
\end{figure}

Fig.~\ref{fig_twinlight_web_interface} shows the preliminary web interface, which shows all relevant aspects of the \ac{DT}.
One of the tabs shows the loaded topology with pan and zoom controls, and selecting a link displays the associated devices and their respective physical-layer parameter values.
Another tab shows all currently established services.
Selecting a specific service highlights its hop-by-hop path, along with the \ac{ROADM} and \ac{EDFA} hinges that contribute to its \ac{PDL} penalty, thereby making the hinge model visible.
From a service, you can navigate to its monitoring page.

The monitoring page shows a current-value card with a sparkline and a historical chart for each of the six \ac{OPM} metrics, alongside the constellation and eye diagrams synthesized from the same physics state.
The three views together expose the perturbation of \ac{QoT} at progressively finer levels of abstraction, from aggregated metrics through scatter statistics to time-domain waveforms.

The web interface also exposes a tab that allows the audience to request a lightpath between two nodes.
The user can select a modulation format or leave it as the best available for the path.
After adding a new lightpath, the audience can then inspect its real-time \ac{OPM} metrics.

Finally, the web interface also contains tabs that allow the audience to inspect the network inventory.
This includes a complete list of equipment, a page that allows the user to compute the $k$-shortest paths between two nodes, and one that shows the spectrum allocation for all links in the network.

\section{Conclusions}

TwinLight unifies a GNPy steady-state baseline with four analytical transient models via standards-compliant \ac{T-API} REST and telemetry-streaming interfaces.
By behaving as a complete \ac{ODC} to any conformant orchestrator, the tool is positioned to accelerate the validation of \ac{SDN} control logic, the training and testing of agentic and \ac{ML}-driven optical-network algorithms, and the empirical study of physical-layer dynamics under realistic provisioning loads.
We release TwinLight as open source so that the community can adopt, extend, and integrate it into pipelines that today rely on laboratory hardware, typically subject to production-readiness validation.
Open extensions include multi-band physics, telemetry streaming coverage for connectivity-service objects, and a fault-prediction pipeline backed by the same synthesized observability.

%-------------------------------------------------- Acknowledgements Section -------------------------------------------------------%
\clearpage
\section{Acknowledgements}
This work has been supported by the Horizon Europe ECO-eNET project, funded by the SNS JU under grant agreement No. 101139133. %, and by....

%-------------------------------------------------- Bibliography Section -------------------------------------------------------%
% see also https://tex.stackexchange.com/questions/55030/text-before-references-but-after-bibliography-title-with-bibtex as of 2024-02-29
\printbibliography

\vspace{-4mm}

%%%%%%%%%%%%%%%%%%%%%%%%%%%%%%%%%%%%%%%%%%%%%
%---------------------------------------------- End of Document -----------------------------------------------%
\end{document}